\documentclass[aps,prx,a4paper,twocolumn,superscriptaddress,citeautoscript,longbibliography]{revtex4-2}
\usepackage{soul}
\usepackage[normalem]{ulem}
\usepackage{hyperref}
\usepackage[LGR,T1]{fontenc}
\usepackage{textcomp}
\usepackage{amssymb}
\usepackage{graphicx}
\usepackage{subscript}
\usepackage{amsmath}
\usepackage{ulem}
\usepackage{appendix}
\usepackage{xcolor}

\makeatletter

\DeclareFontEncoding{LGR}{}{}
\DeclareTextSymbol{\~}{LGR}{126}

\makeatother

\usepackage{babel}
\begin{document}

\title{The spectral weight of hole doped cuprates across the pseudogap critical point}

\author{B.~Michon}
\affiliation{Department of Quantum Matter Physics, University of Geneva, 24  Quai Ernest-Ansermet, 1211 Geneva 4, Switzerland}
\author{A.B.~Kuzmenko}
\affiliation{Department of Quantum Matter Physics, University of Geneva, 24  Quai Ernest-Ansermet, 1211 Geneva 4, Switzerland}
\author{M.K.~Tran}
\affiliation{Department of Quantum Matter Physics, University of Geneva, 24  Quai Ernest-Ansermet, 1211 Geneva 4, Switzerland}
\author{B.~McElfresh}
\affiliation{Department of Quantum Matter Physics, University of Geneva, 24  Quai Ernest-Ansermet, 1211 Geneva 4, Switzerland}
\author{S.~Komiya}
\affiliation{Central Research Institute of Electric Power Industry, Materials Science Research Laboratory, 2-6-1 Nagasaka, Yokosuka, Kanagawa, Japan}
\author{S.~Ono}
\affiliation{Central Research Institute of Electric Power Industry, Materials Science Research Laboratory, 2-6-1 Nagasaka, Yokosuka, Kanagawa, Japan}
\author{S.~Uchida}
\affiliation{Department of Superconductivity, University of Tokyo, Yayoi 2-11-16, Bunkyo-ku, Tokyo 113, Japan}
\author{D.~van~der~Marel}\email{dirk.vandermarel@unige.ch}
\affiliation{Department of Quantum Matter Physics, University of Geneva, 24  Quai Ernest-Ansermet, 1211 Geneva 4, Switzerland}

\begin{abstract}
\noindent
One of the most widely discussed features of the cuprate high $T_c$ superconductors is the presence of a pseudogap in the normal state~\cite{keimer2015,vishik2018}. Recent transport and specific heat measurements have revealed an abrupt transition at the pseudogap critical point, denoted $p^*$, characterized by a drop in carrier density~\cite{badoux2016,collignon2017,michon2018} and a strong mass enhancement~\cite{michon2019,girod2021}. In order to give more details about this transition at $p^*$, we performed low-temperature infrared spectroscopy in the normal state of cuprate superconductors La$_{2-x}$Sr$_x$CuO$_4$ (LSCO) and La$_{1.8-x}$Eu$_{0.2}$Sr$_x$CuO$_4$ (Eu-LSCO) for doping contents across the pseudogap critical point $p^*$ (from $p = 0.12$ to $0.24$). Through the complex optical conductivity $\sigma$, we can extract the spectral weight, $K^*$, of the narrow Drude peak due the coherent motion of the quasi-particles, and the spectral weight enclosed inside the mid-infrared (MIR) band, $K_{MIR}$, caused by coupling of the quasi-particles to collective excitations of the many-body system. $K^*$ is smaller than a third of the value predicted by band calculations, and $K_{MIR}$ forms a dome as a function of doping. We observe a smooth doping dependence of $K^*$ through $p^*$, and demonstrate that this is consistent with the observed doping dependence of the carrier density and the mass enhancement. We argue that the superconducting dome is the result of the confluence of two opposite trends, namely the increase of the density of the quasi-particles and the decrease of their coupling to the collective excitations as a function of doping.
\end{abstract}

\pacs{74.25.F, 74.45.+c, 74.70.Tx}

\maketitle

\section{Introduction}

\begin{figure*}[htb]
\begin{center}
\includegraphics[width=15cm]{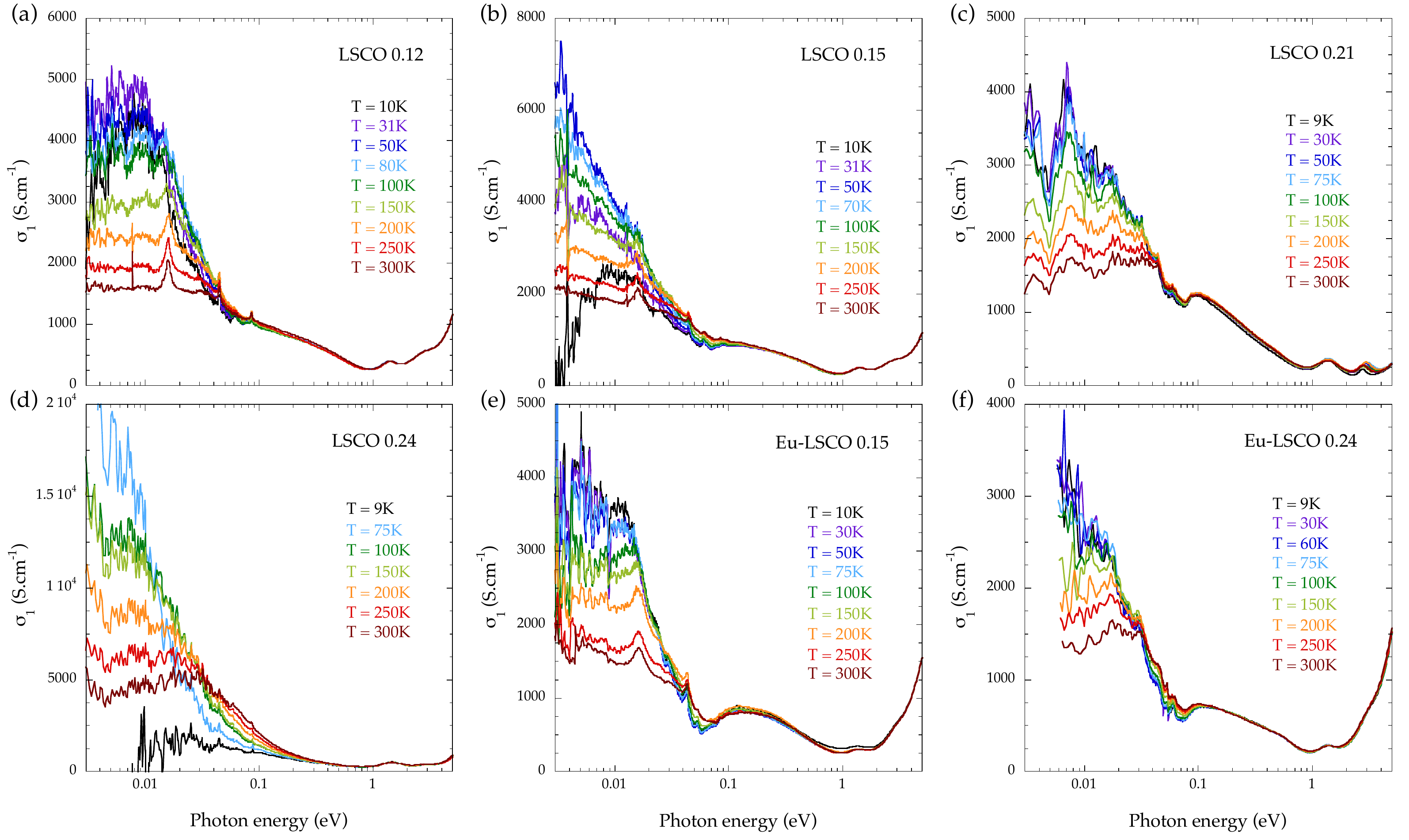}
\caption{ Real part of the $ab$-plane optical conductivity, $\sigma_1(\omega)$ of La$_{2-x-y}$Eu$_{y}$Sr$_x$CuO$_4$ single crystals ($x=0.12,0.15,0.21,0.24$,  $y=0.0,0.2$). 
 \label{fig:1}}
\end{center}
\end{figure*}

The rich phase diagram of hole doped cuprates is a major challenge in condensed matter physics. In addition to superconductivity with the highest superconducting $T_c$ at ambient pressure ($T_c\sim 100$~K) it contains several other competing phases including the pseudogap that is most prominent for very low carrier concentrations.  While many proposals have been made about the nature of the pseudogap phase, the microscopic description thereof is not universally agreed upon~\cite{keimer2015,vishik2018}. Moreover, the superconducting pairing mechanism as well as its link to the pseudogap phase has not been established beyond any doubt. When the carrier density exceeds the value where the highest $T_c$ is observed (the so-called “optimal doping”), a critical point $p^*$ is reached where the pseudogap vanishes. At this critical doping $p^*$ two important signatures have been observed: firstly the carrier density $n$ observed by Hall effect~\cite{badoux2016,collignon2017,putzke2021} and thermal transport~\cite{michon2018} jumps from $p$ to $1+p$ indicating a major Fermi surface reconstruction, secondly the density of states and associated effective mass $m^*$ observed by NMR~\cite{kawasaki2010} and specific heat~\cite{michon2019,girod2021} is strongly enhanced at $p^*$ and exhibits a logarithmic divergence in temperature.
The jump in $n$ and the strong renormalization of $m^*$ at the critical doping $p^*$ underline that the pseudogap phase terminates at this point by a quantum phase transition. 
The close proximity of $p^*$ to optimal doping further suggests a key role in the mechanism of superconductivity in the cuprates~\cite{she2011}. 

Two additional ordering phenomena occur at a doping level close to the point where the pseudogap vanishes: (i) A change of crystal structure from tetragonal at high doping to orthorhombic (approximately a $\sqrt{2}\times\sqrt{2}$ superstructure) at low doping~\cite{jorgensen1988,keimer1992,klauss2000}. (ii) A Lifshitz transition of the Fermi surface topology: at low doping it is a hole-like pocket centered at the Brillouin zone corner, at high doping it is an electron-like pocket around the Brillouin zone center~\cite{park2014,horio2018,miao2021}. For La$_{2-x}$Sr$_x$CuO$_4$ (LSCO) both the structural transition~\cite{keimer1992} and the Lifshitz transition~\cite{horio2018} have been situated at $p=0.20$. For La$_{1.8-x}$Eu$_{0.2}$Sr$_x$CuO$_4$ (Eu-LSCO) the structural phase diagram is more complex due to three different structural variants~\cite{klauss2000}, but the electronic specific heat is strongly peaked at $p=0.235$~\cite{michon2019}. At approximately the same doping also the orthorhombic distortion disappears~\cite{klauss2000}. Approximately the same critical doping is observed for the Lifshitz transition, the vanishing of the pseudogap and the structural transition. In the sequel{\color{blue},} we will for each of the two types of compounds of the present study refer to a single critical doping, namely $p^*(LSCO)=0.20$ and $p^*(Eu-LSCO)=0.235$.

Previous optical studies on bismuth and mercury cuprate families~\cite{mirzaei2013,heumen2009a,heumen2009b} have demonstrated that the free carrier spectral weight is far below the band calculations and approaches zero at zero doping. These studies have also demonstrated the coupling to high energy degrees of freedom far exceeding those of phonons supporting the possibility of an electronic pairing mechanism. However, the strongly overdoped side of the phase diagram has remained unexplored and the high $T_c$ of the Hg1201 and Bi2212 families has limited the accessible temperature range where the normal state properties could be analyzed.

To elucidate this puzzle, we measured infrared optics and extracted the complex optical conductivity $\sigma$ in the normal state of two cuprates, LSCO and Eu-LSCO for four dopings across the pseudogap critical point p$^*$: 0.12, 0.15, 0.21 and 0.24 (a total of six samples). In LSCO and Eu-LSCO, the pseudogap line delimited by $T^*$ is well defined both by transport measurements~\cite{cyr2018} and ARPES~\cite{matt2015}. Compared to other cuprates, the $T_c$ of LSCO and Eu-LSCO is relatively low, and thus interesting for extracting the normal state properties down to low temperatures without using external magnetic fields.

The parent compound of LSCO and Eu-LSCO is La$_{2}$CuO$_4$. Substituting La by Sr introduces holes amounting to a hole concentration $p = x$. Substituting La with Eu does not introduce holes or electrons, but it modifies the maximum of the superconducting critical temperature ($T_c^{max}$): in Eu-LSCO $T_c^{max}$ = 15-20~K versus $T_c^{max}$ = 30-40~K in LSCO.
The high temperature crystal structure of LSCO and Eu-LSCO for all Sr concentrations is body centered tetragonal (space group $I4/mmm$) with lattice parameters $a = b =  3.78$ \AA~and $c = 13.2$ \AA. For LSCO with doping $p<p^*(LSCO)$, the low temperature structure is single-face-centered orthorhombic (space group $Abma$) with, for the undoped parent compound, $a^{\prime} = 5.42$ \AA,   $b^{\prime}= 5.34$ \AA,  and $c = 13.1$ \AA~\cite{jorgensen1988,keimer1992,klauss2000}.  The orthorhombic distortion corresponds to a rotation of the CuO$_6$ octahedra such as to buckle the Cu-O squares along the $a^{\prime}$ direction, causing the lattice parameter $a^{\prime}$ to be smaller than $b^{\prime}$~\cite{picket1989}. 
Accordingly, in-plane anisotropy has been observed for the magnetic susceptibility~\cite{lavrov2001}, the DC resistivity~\cite{ando2002} and the optical conductivity~\cite{dumm2003} of untwinned crystals with doping concentrations $p<0.04$. The LSCO and Eu-LSCO crystals used for the present study were prepared as described in Refs.~\onlinecite{nakamura1993,hess2003,michon2019,frachet2020}. Since in the orthorhombic phase{\color{blue},} these crystals were $ab$-plane micro twinned, the properties reported in the present study represent the effective medium average of $a$-axis and $b$-axis properties. X-ray diffraction was used to align and select the surface, which for part of the samples was chosen parallel to the $ab$ plane and for the other ones perpendicular to $ab$ such as to obtain surface dimensions of at least 0.5 mm$^2$. A polarizer was used for selecting the optical response parallel to the CuO$_2$ planes.

\section{Optical conductivity}
We measured the infrared reflectivity from 2~meV to 0.5~eV using a Fourier transform spectrometer with a home design UHV optical flow cryostat and in-situ gold evaporation for calibrating the signal. In the energy range from 0.5~eV to 5~eV we measured real and imaginary parts of the dielectric function using a home-design UHV cryostat installed in a Visible-UV ellipsometer. Combining  the ellipsometry and reflectivity data and using the Kramers-Kronig relations between the reflectivity amplitude and phase, provided for each measured temperature the complex optical conductivity spectrum $\sigma(\omega,T) = \sigma_1(\omega,T) + i\sigma_2(\omega,T)$ in the range from 2~meV to 5~eV.  

The optical conductivity spectra of LSCO and Eu-LSCO hole-doped from $p=0.12$ to $p=0.24$ are presented in Fig.~\ref{fig:1}.
The doping content $p = 0.15$ corresponds to the “optimal” doping where $T_c$ is maximum~\cite{boebinger1996,cooper2009}. In these figures, we clearly observe a broad MIR band located around 0.1-0.3 eV, which is more pronounced in underdoped and optimally doped samples compared to the Drude response. Below $T_c$, a depletion at low energy is present, corresponding to the superconducting gap.

\begin{figure}
\includegraphics[width=6cm]{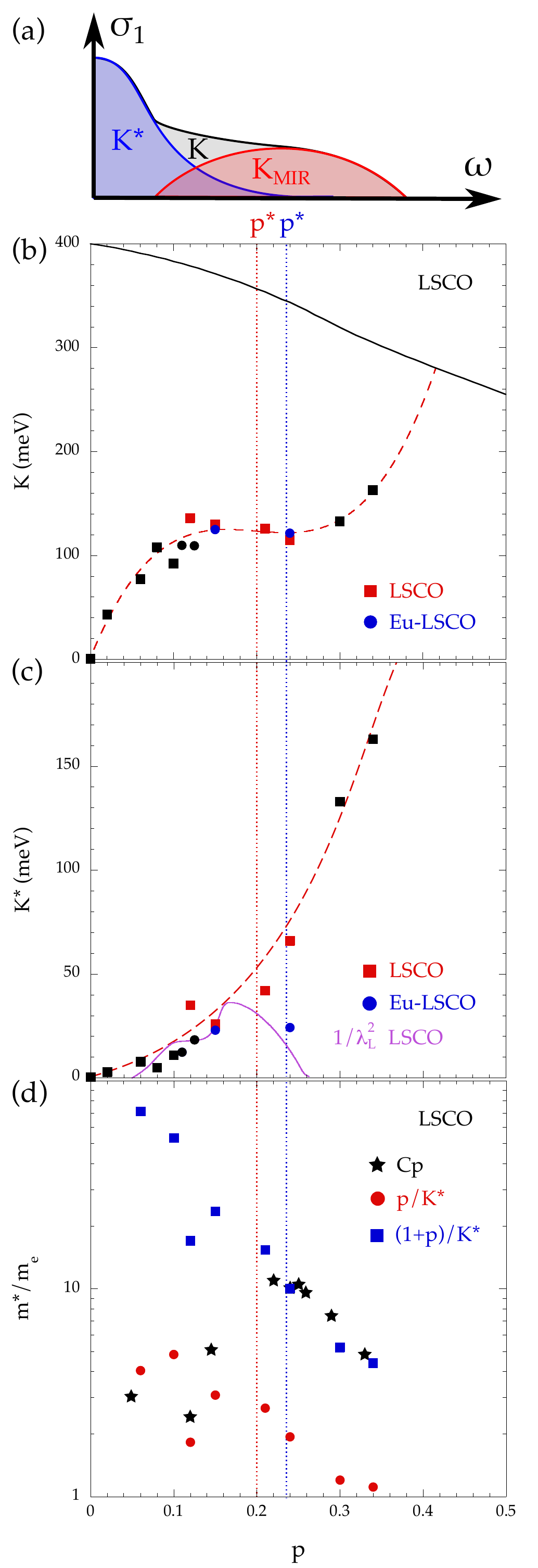}
\caption{
(a) Sketch of the two components of the free carrier optical conductivity: Zero-energy mode with spectral weight $K^*$ and mid-infrared band with spectral weight  $K_{MIR}$.
(b) Total free carrier spectral weight $K=K^*+K_{MIR}$ in a broad range of hole doping for LSCO (red squares) and Eu-LSCO (blue circles). Black squares and black circles are $K$ from previously published optical data~\cite{uchida1991,uchida1996,pignon2008,autore2014}. The solid black curve represents the theoretical value $K_{th}$ extracted from band calculations.
(c) Spectral weight of the zero-energy mode $K^*$. Solid purple line: superconducting spectral weight $K_{sc}^*$ calculated from the London penetration depth~\cite{panagopoulos2003,bozovic2016}. 
(d) Effective mass in units of the free electron mass $m_e$ obtained from the linear term of the specific heat 
(black stars~\cite{girod2021}) compared to the effective mass from $K^*$ assuming a Fermi liquid state ($1+p$ free carriers, blue squares) and a doped Mott insulator ($p$ free carriers, red circles).
The red and blue vertical lines indicate respectively the critical dopings $p^*$ for LSCO and for Eu-LSCO separating the orthorhombic ($p<p^*$) and tetragonal ($p>p^*$) crystallographic phases.
 \label{fig:2}}
\end{figure}
The optical conductivity probes the collective response due {\color{blue}to} the free charge carriers, phonons, and inter-band transitions. For a non-interacting Fermi-gas, the free charge response is described by a zero-energy mode (the Drude peak) with a spectral weight $K$ and the width corresponding to the (frequency independent) scattering rate. If we take into account coupling to phonons or collective excitations of the electron many-body system, part of the Drude spectral weight is transferred to a mid-infrared band (MIR band) at higher energy, carrying a spectral weight $K_{MIR}$. 
These two components, schematically sketched in Fig.~\ref{fig:2}(a), are clearly visible in the experimental spectra shown in Figs.~\ref{fig:1}(a-f). 
The residual Drude response carries a spectral weight $K^*$. The total spectral weight becomes
\begin{equation}
K = K^*+ K_{MIR}.
\end{equation}
A convenient tool for the analysis of the optical conductivity of an interacting electron liquid is provided by the extended Drude model, where both the scattering rate $1/\tau$ and the effective mass ratio $m^*/m$ are frequency dependent functions~\cite{gotze1972,youn2007,nakajima2010}. For two-dimensional materials, like cuprates, the extended Drude model can be written as
\begin{equation}
\sigma(\omega)=\frac{e^2}{d_c\hbar^2}\frac{K}{1/{\tau}(\omega)-i\omega \, m^*(\omega)/m},
\label{eq:GDM}
\end{equation}
where $d_c$ represents the distance between two CuO$_2$ planes, and $m$ is the band mass. In this expression the relation between spectral weight $K$, density $n$ and mass $m$ of the free charge carriers is provided by the expression $K = d_c\hbar^2{n}/{m}$. A slight rearrangement of terms in Eq.~\ref{eq:GDM} provides
\begin{equation}
\sigma(\omega)=\frac{e^2}{d_c\hbar^2}\frac{K m/m^*(\omega)}{1/{\tau^*}(\omega)-i\omega},
\end{equation}
which is a particularly useful expression in the context of a Fermi liquid. In this case, the scattering rate and effective mass of the quasi-particles are given by  $\lim_{\omega\rightarrow 0}1/\tau^*(\omega)$ and $\lim_{\omega\rightarrow 0}m^*(\omega)$ respectively.
The coherent Drude spectral weight $K^*$ can be defined as 
\begin{equation}
K^* =K\frac{m}{m^*}.
\end{equation}
Experimentally, we obtain $K$ from an integration of the real part of the optical conductivity $\sigma_1$
\begin{equation}
\int_{0}^{\omega_c}\sigma_1(\omega)d\omega = \frac{e^2\pi}{2d_{c}\hbar^2}K,
\end{equation}
with $\omega_c$ the cut-off frequency chosen below the interband transitions and above the MIR band (1 eV in the present case).
An equivalent way to obtain $K$, $K^*$ and $K_{MIR}$, providing the same values, is fitting the spectra with a linear superposition of Lorentzians, and extracting the different spectral weights associated to each intra-band transition: $K^*$ for the narrow Drude response and $K_{MIR}$ for the broad MIR band.

\section{Spectral weight analysis}
\begin{figure*}[htb]
\begin{center}
\includegraphics[width=12cm]{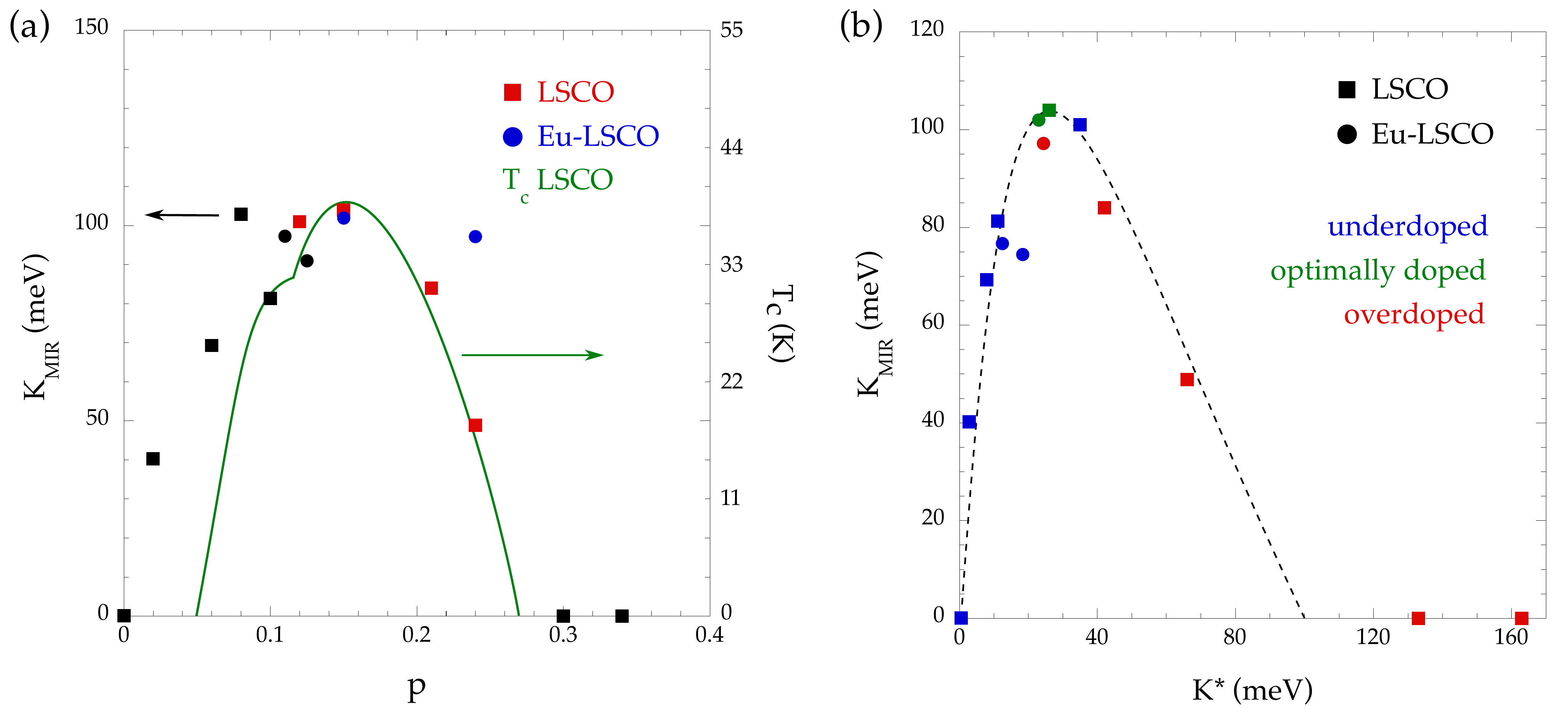}
\caption{
(a) $K_{MIR}$ as a function of p.
(b) $K_{MIR}$ as a function of $K^*$.
Black squares and black circles are from Refs.~\onlinecite{uchida1991,uchida1996,pignon2008,autore2014}, and the $T_c$ dome for LSCO in green line is extracted from Ref.~\onlinecite{takagi1989}.
\label{fig:3}}
\end{center}
\end{figure*}

In the experimental optical conductivities shown in Fig.~\ref{fig:1}{\color{blue},} the two main contributions in the total spectral weight $K$ are from the coherent Drude response $K^*$ and the broad MIR band $K_{MIR}$. Phonons contribute less than 10\% to the total spectral weight $K$.
In Figs.~\ref{fig:2}(b,c) and Fig.~\ref{fig:3}(a), we report the value of $K$, $K^*$ and $K_{MIR}$ as a function of doping. Our data in blue circles (Eu-LSCO) and red squares (LSCO) correspond to the lowest temperature we can reach in the normal state (slightly above $T_c$). Black squares and black circles are values extracted from published data in  low doped  and highly doped samples~\cite{uchida1991,uchida1996,pignon2008,autore2014}.  The red dashed curve~  is a guide to the eye.

Fig.~\ref{fig:2}(b) shows the doping dependence of the total spectral weight $K=K^*+K_{MIR}$. The black curve represents the spectral weight of the sheet conductance obtained from the calculated bandstructure~\cite{pavarini2001} using the expression~\cite{mirzaei2013,norman2007} 
\begin{equation}
K_{th} = \frac{d_c}{V}\sum\limits_{k,\sigma} n_{k,\sigma} \frac{\partial ^2\varepsilon_{k,\sigma}}{\partial k^2}
\label{EqSup:GeneralSpectweight}
\end{equation}
where $n_{k,\sigma}$ is the occupancy of the state with momentum $k$ and spin $\sigma$, $\varepsilon_{k,\sigma}$ is the energy-momentum dispersion, $d_c$ is the interplanar lattice spacing and $V$ is the sample volume.  The bandstructure calculation of Ref.~\onlinecite{pavarini2001} was done for the high temperature tetragonal crystal structure of La$_2$CuO$_4$. Here we assume that the band dispersion is doping independent. These model assumptions do not (and are not intended to) account for  the low temperature orthorhombic structure and/or emerging electronic order in part of the phase diagram. For a two{\color{blue}-}dimensional free electron gas, Eq.~\ref{EqSup:GeneralSpectweight} provides $K_{th} = E_F/\pi$. The effect of interactions is to transfer part of this spectral weight to interband transitions{\color{blue},} which typically spread over a range of 10~eV. An additional effect of electron-electron interactions and electron-phonon coupling is to distribute the free carrier optical conductivity over a “coherent” zero-energy peak and an “incoherent” mid-infrared band. 
For $p<p^*${\color{blue},} the unit cell is doubled and consequently we expect folding of the band structure to occur. This will normally cause a change of the free carrier spectral weight roughly in proportion to the $1+p$ to $p$ reduction of the number of free charge carriers. As stated above, this effect is not taken into account in the bandstructure calculation shown in Fig.~\ref{fig:2}(b). 
A recent {\em ab initio} band structure calculation~\cite{lane2018} has shown that, while antiferromagnetic correlations have a strong effect on the band dispersion, the orthorhombic distortion causes band splittings of order 0.1~eV. The spectral weight removed from the $K^*$ then reappears in the interband transitions in the mid-infrared range. For this reason{\color{blue},} the calculated $K_{th}$ shown in Fig.~\ref{fig:2}(b) 
should be compared to the experimental $K=K^*+K_{MIR}$ ({\em i.e.} instead of $K^*$).

For $p<p^*(LSCO)${\color{blue},}  the experimental value of $K$ initially increases roughly proportional to the hole doping $p$. Then, between $p = 0.10$ and $0.30$, $K$ remains at an almost constant value of about 125 meV, which falls a factor 3 below $K_{th}$. 
This significant reduction of the free carrier spectral weight can be explained by the fact $K \propto n/m$, where $n$ is small inside the pseudogap state.  
Above $p = 0.30$, $K$ finally increases again, and the extrapolation of the experimental data reaches the band value $K_{th}$ around $p = 0.40$.

In Fig.~\ref{fig:2}(c) the narrow Drude spectral weight $K^*$ is presented as a function of doping. Similar as for $K$, $K^*=0$ in the insulating phase and it increases gradually when the system gains more and more holes. It increases continuously, approximately as a quadratic function of doping, up to $p = 0.34$, without any noticeable anomaly ({\em i.e.}~a kink or a jump) at $p^*$. In a free electron approximation where $K^*\propto n/m^*${\color{blue},} our observations for $K^*$ would be in conflict with the sharp transition in carrier density observed in the DC transport data~\cite{badoux2016,collignon2017,michon2018}. However, for an interacting Fermi liquid, the Drude spectral weight $K^*$ is obtained by replacing $\varepsilon_{k,\sigma}$ in  Eq.~\ref{EqSup:GeneralSpectweight} with the quasiparticle energy $\varepsilon^*_{k,\sigma }$. This expression mixes two different quantities, $n_{k,\sigma}$ and $1/m_{k,\sigma}^*=\hbar^{-2}\partial ^2\varepsilon^*_{k,\sigma}/\partial k^2$, both of which undergo important changes around $p^*$. A detailed description of the doping dependence of $K^*$ requires {\em ad hoc} assumptions about the evolution of the electronic structure in the pseudogap phase. Instead, staying close to the experimentally observed quantities, we present in Fig.~\ref{fig:2}(d) a comparison between the effective mass obtained from  the linear term of the specific heat $\gamma$~\cite{girod2021} using the relation $m^*_{sh} = 3 d_c \hbar^2\gamma/(\pi k_B^2)$, and the effective mass from the spectral weight of the zero-energy mode assuming a Fermi liquid using $m^*_{F}=d_c\hbar^2 (1+p)/K^*$ (blue squares) and a doped Mott insulator using $m^*_{M}=d_c\hbar^2 p/K^*$ (red circles).
We see that $m^*_{sh}$ matches $m^*_{M}$  for  $p < p^*(LSCO)$, and $m^*_{F}$ for  $p > p^*(LSCO)$. This comparison puts in evidence a transition in the carrier density around $p = p^*(LSCO)$, in agreement with the transition observed by transport measurements.

In a recent study of LSCO~\cite{horio2018}, the linear term in the specific heat, $\gamma$, which is proportional to the density of states at $E_F$, was compared to the value expected from the electronic band structure $\varepsilon_{k,\sigma}$ measured with angle resolved photoelectron spectroscopy (ARPES), with the result that $\varepsilon_{k,\sigma}$ measured with ARPES gives a $\gamma$ in overdoped LSCO well below the specific heat value. It was speculated that the mass enhancement of $\gamma$ in the electronic specific heat results from quantum criticality emerging from the pseudogap collapse. The good agreement Fig.~\ref{fig:2}(d) between the optical data and specific heat on the overdoped side would suggest that the renormalization of the spectral function near the critical doping $p=p^*(Eu-LSCO)$, is more easily resolved with optics than with ARPES.

For comparison{\color{blue},} the penetration depth $1/\lambda_L^2$ is plotted in Fig.~\ref{fig:2}(c) as a solid purple line~\cite{panagopoulos2003,bozovic2016}. As $1/\lambda_L^2 \propto n/m^*$ inside the superconducting state, we can calculate the superconducting spectral weight $K_{sc}^*$ of the condensed Cooper pairs. From $p = 0$ to the optimally doped sample $p = 0.15$, $K_{sc}^* = K^*$, which is what we should expect if the pair breaking gap $\Delta$ is much bigger than the residual scattering rate $\hbar/\tau_0$. This observation does not in itself allow to distinguish between the BCS limit where $\Delta$ is much smaller than the Fermi energy $E_F$, or the opposite case corresponding to Bose-Einstein condensation of preformed pairs~\cite{uemura1988,randeria1989,marel1992,rietveld1992,alexandrov1994}. On the overdoped side, the observation that $K_{sc}^*$ goes to zero while $K^*$ keeps increasing, indicates that $\Delta<\hbar/\tau_0$. This is a natural consequence of $\Delta$ becoming smaller for reasons that we will discuss below. 

By knowing $K$ and $K^*$, we can directly extract $K_{MIR} = K -K^*$, the spectral weight enclosed inside the MIR band corresponding to the electronic correlations in the extended Drude model. We show $K_{MIR}$ as a function of doping in Fig.~\ref{fig:3}(a) and compare it with $T_c$ (solid green curve~\cite{takagi1989}). 
We give another representation of this behaviour in Fig.~\ref{fig:3}(b), where $K_{MIR}$ is plotted as a function of $K^*$, which is a monotonously increasing function of doping. The three colors represent the different doping regions: underdoped in blue, optimally doped in green and overdoped in red. It is interesting to note that this representation results in a smoother doping dependence, indicating that any scatter in the doping value obtained from sample stoichiometry (for example oxygen off-stoichiometry) is absent when we use $K^*$ as a measure of the charge carrier density. 
We observe that the MIR spectral weight forms a dome with a maximum around $p = 0.15$, {\em i.e.} $K^*=26$~meV. This doping value $p = 0.15$ matches the optimal doping for $T_c$ obtained in Refs.~\onlinecite{panagopoulos2003,bozovic2016}. This correlation of the MIR spectral weight and $T_c$ as a function of doping suggests that the strength of the MIR band reveals an important part of the superconducting pairing interaction.

In Ref.~\onlinecite{mirzaei2013}{\color{blue},} a linear relation between $p$ and $K^*$, $K^* = K_0 p$ was reported for HgBa$_2$CuO$_{4+\delta}$ (Hg1201), Bi$_2$Sr$_2$CaCu$_2$O$_{8-\delta}$ (Bi-2212) and Bi$_2$Sr$_2$CuO$_{6+\delta}$ (Bi-2201) with $K_0=0.5$~eV. In contrast{\color{blue},} the results for LSCO and Eu-LSCO shown in Fig.~\ref{fig:2}(c) follow a superlinear dependence. The aforementioned linear relation would give $K^*\approx 0.2$~eV at $p=p^*(LSCO)$, which is two times higher than the value for LSCO shown in Fig~\ref{fig:2}(c). Given that the $ab$ planes of LSCO are buckled whereas Hg1201, Bi-2212, and Bi-2201 have flat $ab$ planes, we postulate that the buckling of the $ab$ planes of LSCO causes a further suppression of $K^*$ in addition to the effect of the pseudogap that is common to all cuprate high $T_c$ superconductors. 

\section{$T_c$ and the spectral weight}
\begin{figure*}[ht!]
\begin{center}
\includegraphics[width=12cm]{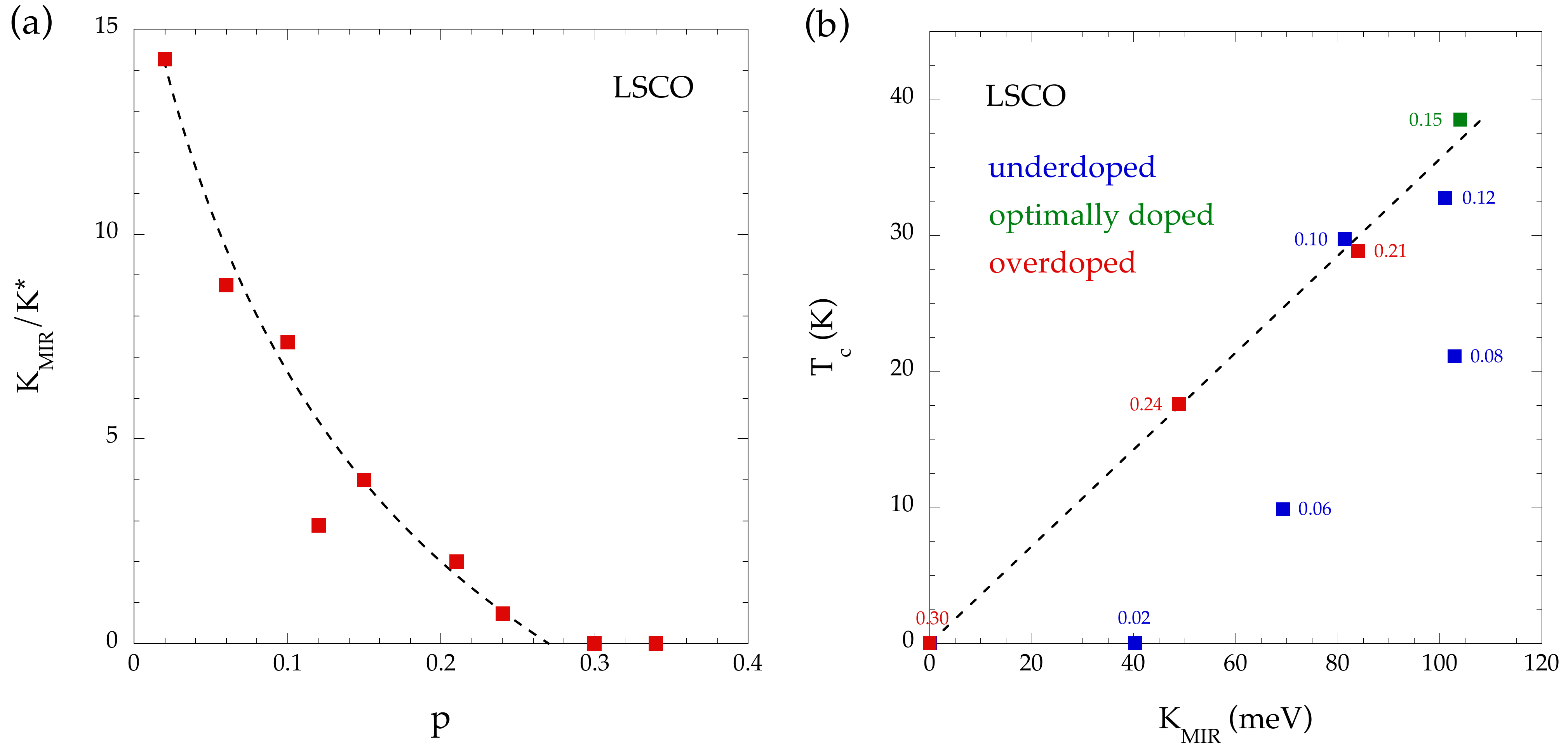}
\caption{
(a) Spectral weight ratio $K_{MIR}/K^*$ as a function of hole doping - (b) $T_c$ as a function of $K_{MIR}$. The spectral weight ratio  $K_{MIR}/K^*$ vanishes around $p = 0.27$ exactly where the superconducting dome ends. 
\label{fig:4}}
\end{center}
\end{figure*}
For the overdoped samples where no band folding is present ($n=1+p$), the ratio $K_{MIR}/K^*$ describes the mass enhancement $\lambda = m^*(0)/m -1= K_{MIR}/K^*$ due to the coupling to electronic and vibrational collective excitations. At the underdoped side of the phase diagram where $n=p$, $K^*$ is reduced due to two effects, namely the reconstruction of the Fermi surface due to band folding, as well as mass enhancement due to many-body effects. In Fig.~\ref{fig:4}(a), $K_{MIR}/K^*$ is represented as a function of doping content in red squares. The black dashed line is a guide to the eye. In the underdoped region, $K_{MIR}/K^*$  is very high and goes to infinity down to $p = 0$. This is consistent with the notion that the charge in a Mott-insulator is fully localized, hence bound at a non-zero energy. In the overdoped region, $K_{MIR}/K^*$ becomes indistinguishable from zero within the measurement accuracy for $p \ge 0.27$, coinciding with the upper doping limit of the superconducting dome.
In Fig.~\ref{fig:3}(a), the $T_c$-like dome of $K_{MIR}$ puts in evidence a potential link between $K_{MIR}$ and $T_c$. For this reason, in Fig.~\ref{fig:4}(b), we plot $T_c$ as a function of $K_{MIR}$. Underdoped samples are in blue squares, optimally doped in green square and overdoped samples in red squares.  For $p\ge 0.10${\color{blue},} we find a proportionality between these two quantities (with the familiar exception at $1/8$ doping), which implies that the broad MIR band not only results from coupling of the electrons to collective (spin, charge and vibrational) excitations, but also that this coupling mediates superconducting pairing. For the strongly underdoped samples ($p\le 0.08$) $T_c(p)$ is much lower, indicating that in the underdoped region the coupling to collective excitations contributes only in part to the superconducting pair formation. 

\section{Conclusions}
In this study, the optical conductivity $\sigma(\omega,T)$ was measured in the normal state of La$_{2-x}$Sr$_x$CuO$_4$ and La$_{1.8-x}$Eu$_{0.2}$Sr$_x$CuO$_4$ at four doping contents $p = 0.12, 0.15, 0.21$ and $0.24$ across the pseudogap critical point $p^*$.
The free carrier spectral weight exhibits the following trends as a function of doping: (i) the total spectral weight $K$ is much smaller than the band calculation, illustrating the consequences of the pseudogap state and the strong electronic correlations in the underdoped region. (ii) The coherent spectral weight $K^*$ increases continuously and smoothly from $p = 0$ to $0.34$ without any noticeable transition.
Due to a compensating effect in $K^*$, which is proportional to the ratio of carrier density over effective mass, this is consistent with anomalies near p$^*$ observed in the specific heat and the Hall constant.
(iii) The spectral weight of the broad MIR band forms a dome looking alike the superconducting dome and T$_c\propto K_{MIR}$, putting in evidence a candidate for the pairing mechanism revealed by the MIR band. In particular, $T_c$ goes to zero at the same doping value where the intensity of the MIR band becomes negligible. (iv)  $K^*_{sc} = K^*$ in the underdoped region and $K^*_{sc} \ll K^*$ in the overdoped region.

Taken together{\color{blue},} the superconducting dome in the phase diagram of LSCO and Eu-LSCO cuprates is the result of two competing phenomema, namely (i) the  density of charge carriers available for superconductivity characterized by $K^*$, which vanishes for doping $p\rightarrow 0$ and (ii) the pairing interaction revealed by $K_{MIR}$, which drops to zero at high doping.

The datasets generated and analyzed during the cur-rent study are available in Ref.~\onlinecite{yareta2021} and will be preserved for 10 years.

\section*{Acknowledgments}
We thank Antoine Georges and Christophe Berthod for fruitful discussions, and Bernd B\"uchner, Sunseng Pyon, Tomohiro Takayama, and Hidenori Takagi for providing samples. This project was supported by the Swiss National Science Foundation through project 200020-179157. 
%
%
%
%
\end{document}